\documentclass[prl,twocolumn,groupedaddress,showpacs,floatfix]{revtex4}
\usepackage{graphicx}
\begin{document}
\bibliographystyle{apsrev}
\title{Full counting statistics of multiple Andreev reflection}
\date{\today}
\author{G\"oran Johansson}
\affiliation{Institut f\"ur Theoretische Festk\"orperphysik, 
Universit\"at Karlsruhe, D-761 28 Karlsruhe, Germany}
\author{Peter Samuelsson}
\affiliation{D\'epartement de Physique Th\'eorique, Universit\'e de
Gen\`eve, CH-1211 Gen\`eve 4, Switzerland.}  
\author{\AA ke Ingerman}
\affiliation{Department of Physics, University of the Western Cape,
Cape Town, South Africa}


\begin{abstract}
We derive the full counting statistics of charge transfer through a
voltage biased superconducting junction. We find that for measurement
times much longer than the inverse Josephson frequency, the counting
statistics describes a correlated transfer of quanta of multiple
electron charges, each quantum associated with the transfer of a
single quasiparticle. An expression for the the counting statistics in
terms of the quasiparticle scattering amplitudes is derived.
\end{abstract}
\maketitle

Due to the discrete nature of electric charge, the current in
mesoscopic conductors generally fluctuates. Over the last decade,
there has been an increasing interest, theoretical as well as
experimental \cite{Buttrev}, in the physics of current
fluctuations. Most studies have been focused on noise, the second
moment of the fluctuations, but recently a considerable interest has
been shown for the full distribution of charge fluctuations, the full
counting statistics (FCS) \cite{Levitov1}. A variety of theoretical
approaches to the FCS, ranging from quantum
mechanical \cite{Levitov2,Beenakker}, via quasiclassical
\cite{Nazarov} to classical \cite{Pilgram}, have been developed. The
third moment of current fluctuations was very recently measured
\cite{Reulet}, opening up the road, as well, to experimental
investigation of the higher moments of the fluctuations.

Noninteracting electrons in purely normal conductors are transferred
one by one \cite{Levitov1}. In normal-superconducting junctions, the
charge transfer mechanism across the normal-superconducting interface,
at energies below the superconducting gap, is Andreev reflection. As a
consequence, the FCS include terms describing 
correlated transfer of pairs of electrons \cite{MuzKhmel}. 
Recently, Belzig
and Nazarov \cite{Belzig2} studied the FCS in
superconducting junctions with a fixed phase difference between the
superconducting electrodes. They found that the classical
interpretation of the FCS, the probability to transfer a given number
of electrons across the junction during the measurement, could imply
negative probabilities. Coupling the junction to a detector
\cite{Nazarov1}, they showed that this resulted from an attempt to
interpret the phenomena of supercurrent with
classical means.

In voltage biased superconducting junctions, the physical situation is
quite different. Due to the applied voltage bias, the superconducting
phase difference oscillates with the Josephson frequency $2eV/\hbar$,
giving rise to both dc and ac-components of the current.  
For measurement times much longer than the inverse Josephson frequency,
only the dc-current, which is dissipative, contributes to the net
charge-transport. Microscopically, the charge is transported between the two 
superconductors via coherent 
multiple Andreev reflections (MAR) \cite{OldMAR}.
The current has been studied in 
various junctions,  both theoretically \cite{TheoryMAR} and experimentally
\cite{ExpMAR}. Recently, 
also the noise was studied \cite{Noisetheory,Naveh,Noiseexp}.

In this paper we present the FCS of charge transfer through a 
voltage biased junction, in terms of the amplitudes
for quasiparticle scattering. Each quasiparticle
scattering process results in an integer number of electron
charges being transferred across the junction.
As a consequence the FCS can be interpreted in classical terms.
At temperatures much lower than the superconducting energy
gap $\Delta$, many-quasiparticle scattering processes are
exponentially suppressed, resulting in a simple probability distribution, 
containing only the {\it probabilities} for {\it single} 
quasiparticle scattering.
This distribution reproduces known results for dc-current and 
zero-frequency noise. We discuss in detail the third cumulant
for single channel junctions and diffusive junctions shorter
than the superconducting coherence length.

\begin{figure}[h]        
\centerline{\includegraphics[width=8cm]{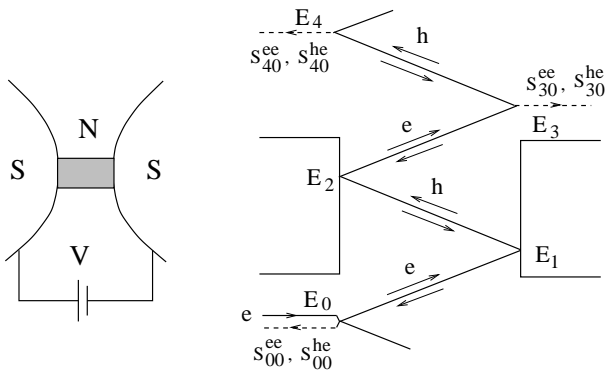}}
\caption{Left: A schematic picture of the junction. 
Right: Multiple Andreev reflection processes
in the voltage biased superconducting junction. An electron-like
quasiparticle injected at an energy $E_0$ can undergo multiple Andreev
reflections before being emitted as an $\alpha$-type (e or h)
quasiparticle at an energy $E_n=E_0+neV$. These scattering processes
have the amplitudes $s^{\alpha e}_{n0}$, shown in the figure. 
Notethat all quasiparticles injected from the left at energies $E_0+2neV$ 
and from the right at $E_0+(2n+1)eV$ scatter on the same ``ladder'' 
in energy space.}
\label{fig1}
\vspace{-.6cm}
\end{figure}

We consider a superconducting junction consisting of two
superconducting reservoirs connected via a normal, mesoscopic
conductor (see Fig. \ref{fig1}). For simplicity of notation, we
consider a junction with a single transport mode, the multi-mode
generalization is discussed below. A voltage $V$ is applied between
the two reservoirs.

The single-particle wavefunctions in the junction, 
solutions to the time dependent
Bogoliubov-de Gennes equation, are scattering states labelled by the
incoming quasiparticle type, injection energy and reservoir. The
scattering states are superpositions of amplitudes for quasiparticles
at energies $\pm neV$ from the injection energy, counted from the
local chemical potential in each contact. The amplitude for an
incoming quasiparticle of type $\alpha$ at energy $E_m$ to exit the
junction as a quasiparticle of type $\beta$ at energy
$E_n=E_m+(n-m)eV$ is denoted $s^{\beta\alpha}_{nm}$ (see
Fig. \ref{fig1}). This amplitude is a function of the scattering
matrix of the normal conductor and the Andreev reflection amplitudes
\cite{Johansson}.

To access the FCS, we make the first important observation that {\it
quasiparticle} scattering in a voltage biased superconducting
junction is formally identical to scattering in a normal
voltage biased junction, with an applied harmonic ac-field. The FCS in such a
system was investigated in detail by Ivanov and Levitov[4]. Following
Ref. [4] we note that for measurement times much longer than the
quasiparticle scattering time, the inelastic single mode scattering
problem can be mapped onto an elastic scattering problem with many
modes. 
The scattering between all energies  and
quasiparticle types of a ``ladder'' (see Fig. \ref{fig1}) is correlated,
while different ladders contribute incoherently.
We denote the ladder by its energy $E_0$ in the interval 
$-\Delta-2eV \leq E_0 < -\Delta$ counting its ``leg'' in the left lead
(e.g. $E_0$ in Fig~\ref{fig1}). 
Thus we may concentrate on the scattering matrix $S$, with elements
 $s^{\alpha\beta}_{nm}$, of a single
ladder, and then integrate over the ladder energy $E_0$.
In general $S$ has infinite dimensions, but in our case the
vanishing probability of Andreev reflection
far outside the gap naturally cuts the number of
relevant modes to the order of $\Delta/eV$.
Quasiparticle current (but not charge current) is
conserved in the scattering processes at the normal-superconductor
interfaces. As a consequence $S$ is
unitary \cite{comment}. 

We are then, in line with \cite{Levitov2,LevitovTD}, 
able to directly write down the
characteristic function in terms of all different many-particle
scattering probabilities
\begin{equation}
\label{FCSdef}
\chi_{E_0}^{qp}({\Lambda})=
\sum_{{\bf{i}},{\bf{o}}} e^{i\left(
\sum_{n\beta\in{\bf{o}}}\lambda_{n\beta}-
\sum_{m\alpha\in{\bf{i}}}\lambda_{m\alpha}\right)}
P_{{\bf{i}}|{\bf{o}}},
\end{equation}
where ${\Lambda}$ is the set of counting fields $\lambda_{n\alpha}$,
one for each mode $n\alpha$.
The outer sum runs over all possible sets of incoming modes
${\bf{i}}=\{m_1\alpha_1, m_2\alpha_2,\dots\}$ and outgoing modes
${\bf{o}}=\{n_1\beta_1, n_2\beta_2,\dots\}$.

The many-particle scattering probabilities are given by
\begin{equation}
\label{NscProb}
P_{{\bf{i}}|{\bf{o}}}=
|s_{\bf{i}}^{\bf{o}}|^2 
\prod_{m\in{\bf{i}}} f(E_m)
\prod_{m'\notin{\bf{i}}} [1-f(E_{m'})],
\end{equation}
where $f(E)=(1+\mbox{exp}[E/kT])$ is the Fermi distribution function
and $|s_{\bf{i}}^{\bf{o}}|$ is the determinant of the matrix
formed by taking the columns $\bf{i}$ and rows $\bf{o}$ of $S$.

Eqs.~(\ref{FCSdef}) and (\ref{NscProb}) gives us the FCS for 
{\it quasiparticle} transfer in a
voltage biased superconducting junction. The object of main interest
is however the FCS of the {\it charge} transfer. We then make the
second important observation that for measuring times much longer than
the inverse Josephson frequency, the net charge transfer is directly
related to the quasiparticle transfer. A quasiparticle of type $\alpha$
incident at energy $E_m$, which is scattered into an outgoing quasiparticle 
of type $\beta$
at energy $E_n$, transports exactly $m-n$ electrical charges across
the junction (for $m<n$, the transported charge is thus negative for
the bias in Fig. \ref{fig1}). This is independent of
the type of quasiparticles $\alpha$ and $\beta$, and also of the charge
of the quasiparticles. This can be shown by studying
any quasiparticle scattering path (see Fig. \ref{fig1}), 
keeping in mind that the
process of Andreev reflection transfers exactly two electrons across
the normal-superconductor interface, while a normal transmission
transfers exactly one electron.

This can also be seen from energy conservation: An electron 
traversing the normal part of the junction from left to right will absorb
the energy quantum $eV$ from the electric field, while an electron
moving in the opposite direction will emit the quantum $eV$.
The effective number of quanta $eV$ absorbed in a quasiparticle
scattering process scattered thus equals the 
number of electrons transferred from left to right.
We emphasize that this approach
correctly counts all the electrons transferred, including the electrons
entering the superconductor as Cooper pairs at energies within the gap.

We can thus count the transferred electrons by counting the transferred
quasiparticles weighted by the number of electrons transferred in each
quasiparticle scattering event. Thus by chosing the counting fields
$\lambda_{n\alpha}=n\lambda$  in
Eq.~(\ref{FCSdef}) we can directly write
down the characteristic function for {\it charge} transfer
\begin{equation}
\label{MARFCSdef}
\chi_{E_0}(\lambda)=
\sum_{{\bf{i}},{\bf{o}}} e^{i\lambda \left(
\sum_{n\beta\in{\bf{o}}} n -
\sum_{m\alpha\in{\bf{i}}} m \right)}
P_{{\bf{i}}|{\bf{o}}},
\end{equation}

Following Ref \cite{Levitov2}, the characteristic function can be 
written on a determinant form as
\begin{equation}
\label{GeneralFCS}
\chi_{E_0}(\lambda)=
\mbox{det}\left[1-\bar f+\bar fS_{-\lambda}^{-1}S_{\lambda}\right],
\label{fcsscat}
\end{equation}
where the elements of the matrix $S_{\lambda}$ are given by
$(S_{\lambda})_{mn}^{\alpha\beta}=s_{mn}^{\alpha\beta}e^{i\lambda(n-m)/2}$
and the diagonal matrix $\bar f$ has elements $\bar
f_{nn}^{\alpha\alpha}=f(E_n)$. This results provides a general
solution to the temperature and voltage dependence of the full
counting statistics of voltage biased superconducting junctions,
for measurement times much longer than inverse Josephson frequency,
and the quasiparticle scattering time, as defined by
the energy dependence of the scattering matrix 
$\sim \hbar \partial_E \ln |s_{mn}^{\alpha\beta}|$.
Eqs.~(\ref{MARFCSdef}) and (\ref{GeneralFCS}) shows that 
the charge is transferred
in correlated quanta of multiple electron charges.

At low temperatures $kT \ll \Delta$
all quasiparticle states below 
the gap are filled, while all states above the gap are empty.
This simplifies Eq.~(\ref{NscProb}) significantly since it
fixes the set of incoming modes $\bf{i}$ to all modes
below the gap, but still all different sets of outgoing modes
should be considered.

The MAR ladder forms a single mode for transport in energy
space \cite{Johansson}. The scattering amplitudes $s_{nm}^{\beta\alpha}$ 
can be decomposed into amplitudes for 
entering the normal region, $t_{m\alpha}^{m\pm}$, 
propagation along the MAR ladder, $t_{mn}^{\uparrow/\downarrow}$, 
and leaving the normal region,  $t_{n\pm}^{n\alpha}$, as
\begin{equation}
s_{nm}^{\beta\alpha}=t_{m\alpha}^{m+} t_{mn}^{\uparrow} t_{n-}^{n\beta},
\ \ t_{mn}^{\uparrow}=t_{mk}^{\uparrow} t_{kn}^{\uparrow} \ \ (m < k < n),
\end{equation}
and analogously for $E_n < E_m$. Using this decomposition we
find that choosing two or more outgoing
modes above the gap in the set $\bf{o}$ gives $P_{\bf{i}|\bf{o}}=0$,
since the matrix $s_{\bf{i}}^{\bf{o}}$ then contains two or more
parallell rows. In other words, the many-particle scattering
process of two or more quasiparticles passing the gap through the
single mode in energy space is prohibited by the Pauli
exclusion principle.
For the outgoing sets with all outgoing modes below the gap
$|s_{\bf{i}}^{\bf{o}}|$ is evaluated through making $s_{\bf{i}}^{\bf{o}}$ 
unitary by inserting a row and a column with the amplitudes 
$t_{m\alpha}^{m+} t_{mg}^{\uparrow}$, where $E_g$ is the first 
energy above the gap in the ladder. Finally, the sets of one 
outgoing mode above the gap, using
similar manipulations, give the single particle scattering
probabilities. The characteristic function can then be expressed
in {\it single-particle scattering probabilities}
\begin{equation}
\chi_{E_0}(\lambda)=
\sum_{m\leq0,n>0}e^{i(n-m) \lambda}
\sum_{\alpha,\beta}|s_{nm}^{\beta\alpha}|^2.
\label{fcsprob}
\end{equation}
This simple expression is the main result of this paper.

\begin{figure}[h]        
\centerline{
\includegraphics[width=4.5cm]{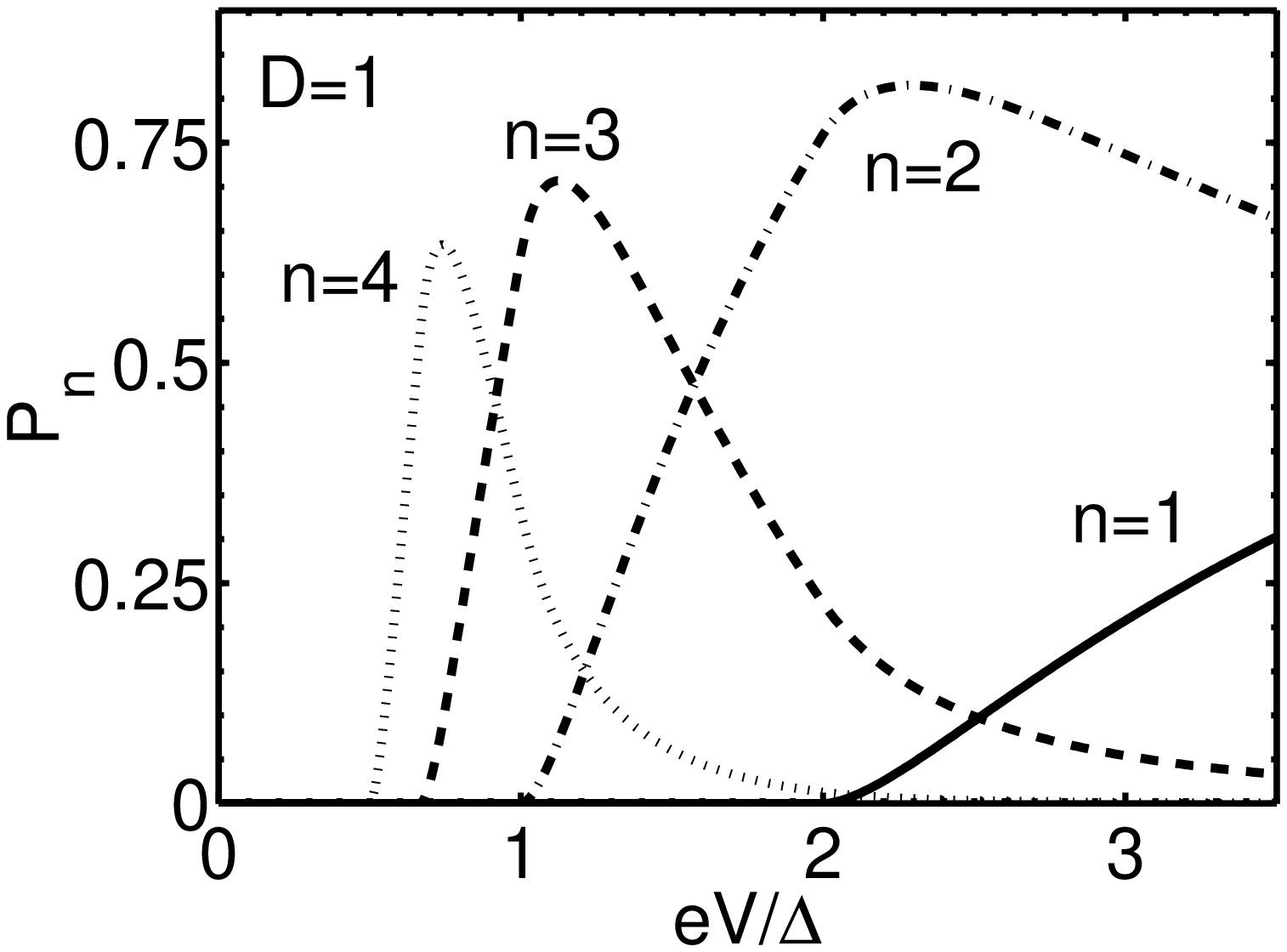}
\includegraphics[width=4.5cm]{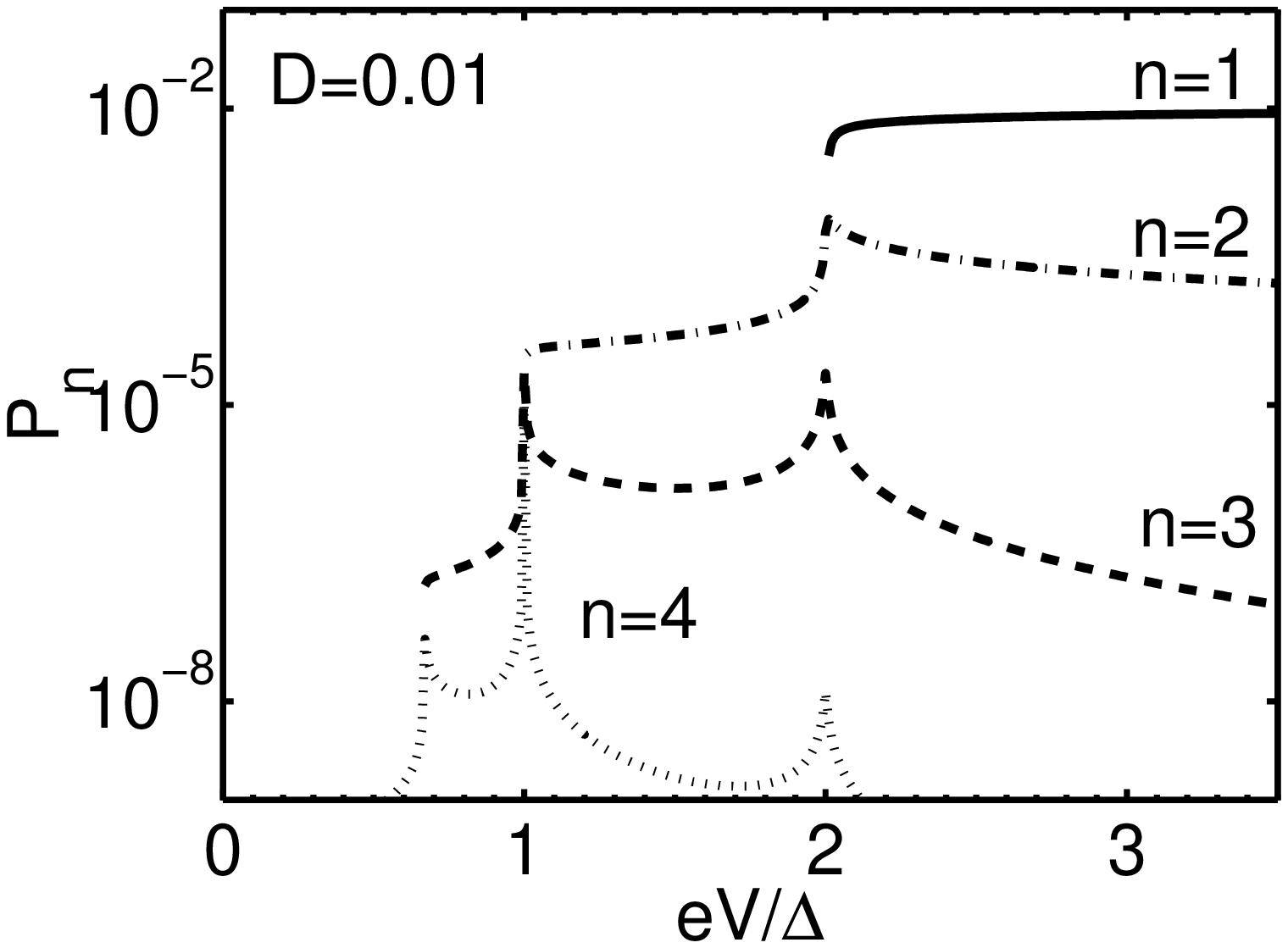}
}
\caption{The probability 
$P_n=1/2eV \int_{-\Delta-2eV}^{-\Delta} P_n(E_0) dE_0$ 
of $n$-electron scattering events, as a function of voltage 
for $n\in\{1,2,3,4\}$ in a fully transparent
junction (left) and a tunnel junction (right). 
}
\label{PnFig}
\end{figure}

The cumulants of the charge transfer distribution function 
are obtained by taking derivatives of $\ln \chi_{E_0}(\lambda)$,
and summing over all ladders (integrating over the energy $E_0$),
\begin{equation}
\label{Ncum}
\langle \langle n(\tau)^m \rangle \rangle=
\frac{\tau}{h}\int_{-\Delta}^{-\Delta-2eV} dE_0
\left. (-i\partial_\lambda)^m \ln \chi_{E_0}(\lambda)\right\vert_{\lambda=0},
\end{equation}
where $\tau$ is the measurement time.

The integrands for the various moments are conveniently expressed in the
physically relevant $n$-electron scattering probabilities \cite{Johansson},
\begin{equation}
P_n(E_0)=
\sum_{m\leq0}\Theta(E_{m+n}-\Delta) \sum_{\alpha,\beta}
|s_{(m+n)m}^{\alpha\beta}(E_0)|^2\ \ \ (n \geq 1),
\end{equation}
with $P_0(E_0)$, the
probability of no quasiparticle passing the gap, giving
the correct normalization $\sum_{n=0}^\infty P_n(E_0)=1$.
In Fig.~\ref{PnFig} the probability of different $n$-electron
scattering processes, as a function of voltage, are shown.

\begin{figure}[h]        
\centerline{
\includegraphics[height=8cm]{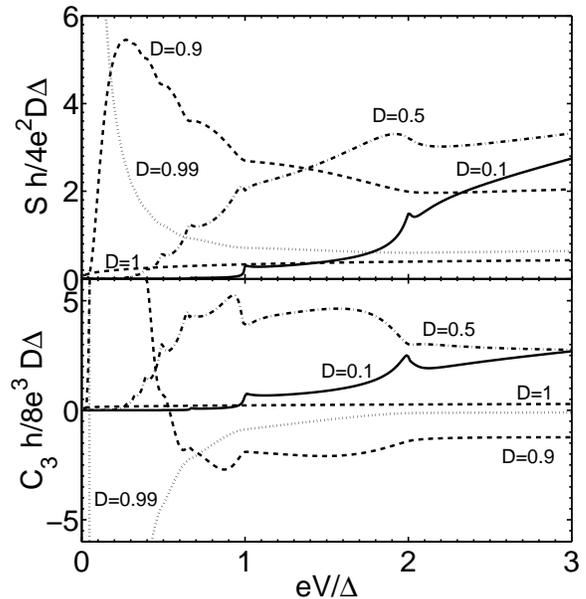}
}
\caption{The zero-frequency current noise (upper panel) and
the third cumulant (lower panel) as a function of voltage for
transparencies $D=0.1,0.5,0.9,0.99$ and $1$. All curves are for a
single mode junction at temperatures $kT \ll \Delta$.}
\label{fig2}
\end{figure}
The integrands for the first three cumulants in Eq.~(\ref{Ncum}) 
are the spectral current density $I(E)=\sum_{n=0}^\infty n P_n(E)$,
the noise spectral density 
$S_I(E)=\sum_{n=0}^\infty n^2 P_n(E) - I(E)^2$, and
the spectral density of the third cumulant
$C_3(E)=\sum_{n=0}^\infty n^3 P_n(E) - I(E)(3S_I(E)+I(E)^2)$.
The expression for the current is identical to the one
presented in Ref.~\cite{Johansson}, and the expression for
the zero-frequency current noise reproduces the known
results \cite{Noisetheory} (see the upper panel in Fig.~\ref{fig2}).

The noise and the third cumulant, calculated numerically, 
are plotted in Fig. \ref{fig2} for different transparencies 
of the normal contact. We see that the subgap structure is 
generally more pronounced in the third cumulant, compared to 
the current and noise. Furthermore $C_3$ changes sign both 
as a function of voltage and transparency. 

\begin{figure}[h]        
\centerline{\includegraphics[width=6.5cm]{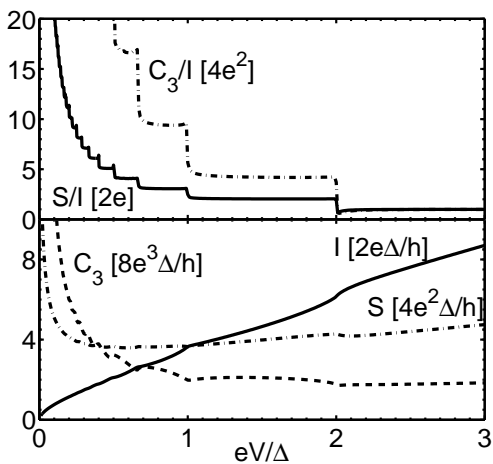}}
\caption{(Upper panel) The Fano factor $S/I$ (solid) and 
$C_3/I$ (dash-dotted) in the tunnel limit $D=0.01$.
(Lower panel) The current (solid), 
noise (dash-dotted) and third cumulant
(dashed) for a short, diffusive superconducting junction.}
\label{fig4}
\end{figure}

In the tunnel limit $D\ll 1$  where all scattering probabilities
in Eq. (\ref{fcsprob}) are small, the corresponding probability
distribution is Poissonian, i.e. describes an uncorrelated
transfer of quantas of multiple charge. Within the $n$-th subgap
region, i.e. $n-1<2\Delta/eV<n$, $n$-electron transfer dominate,
giving $S=2e n I$ and $C_3=(2en)^2 I$, see the upper panel in Fig.~\ref{fig4}.
As seen in the right panel of Fig.~\ref{PnFig} the picture
is more complicated on the borders between the regions.
Due to resonances up to three adjacent $n$-particle processes 
have similar strength \cite{Johansson}, and the FCS is no longer Poissonian.

So far we considered only single mode junctions. 
In the limit of a short
junction, when the scattering matrix of the normal conductor is
independent on energy on the scale of $\Delta$, it is possible, just
as for the current and noise \cite{Naveh}, to write the generating
function in terms of the transmission eigenvalues $D_n$ of the normal
conductor only. For all mesoscopic conductors where the transmission
eigenvalue distribution is known, the FCS can
thus be obtained via averaging the single mode result in
Eq. (\ref{fcsprob}). As an example, we present in the lower panel of
Fig. \ref{fig4} the first three moments for a short, diffusive junction.
We note that the third moment is positive for all voltages. Just as
the noise, the third moment has a divergency for $eV\rightarrow 0$. However,
in an experiment, we expect the fluctuations at small voltages
to be strongly suppressed by inelastic scattering \cite{Gunsenheimer}, 
effectively removing the divergency \cite{Naveh}.

In conclusion, we have derived an expression for the full counting
statistics of a voltage biased superconducting junction, describing 
a correlated transfer of quantas of multiple
electron charge. It is found that the counting statistics can be expressed in
terms of probabilities for quasiparticles to scatter between the two
superconductors.

We acknowledge discussions with Yuriy Makhlin and Dimitri Bagrets. 
This work was supported by
the Humboldt Foundation, the BMBF and the ZIP programme of
the German government (GJ), by the Swiss network MANEP (PS), 
and by the Swedish Foundation STINT (\AA I).


\begin{thebibliography}{99}
\bibitem{comment}
The scattering amplitudes for quasiparticles incoming or
outgoing at energies within the gap, $|E|<\Delta$, are zero.
\end{thebibliography}
\end{document}